\documentclass[conference]{IEEEtran}
\IEEEoverridecommandlockouts
\usepackage{cite}
\usepackage{amsmath,amssymb,amsfonts}
\usepackage{algorithmic}
\usepackage{graphicx}
\usepackage{setspace}
\usepackage{textcomp}
\usepackage{xcolor}
\usepackage{amsmath}
\usepackage{mathtools}
\usepackage{ulem}
\usepackage[font=scriptsize]{caption}
\usepackage{subcaption}

\def\BibTeX{{\rm B\kern-.05em{\sc i\kern-.025em b}\kern-.08em
    T\kern-.1667em\lower.7ex\hbox{E}\kern-.125emX}}
\begin{document}

\title{Bypassing or flying above the obstacles? A novel multi-objective UAV path planning problem\\
{\footnotesize}
\thanks{}
}

\author{\IEEEauthorblockN{Mahmoud Golabi\textsuperscript{1}, Soheila Ghambari\textsuperscript{1}, Julien Lepagnot\textsuperscript{1}, Laetitia Jourdan\textsuperscript{2},\\ Mathieu Br\'evilliers\textsuperscript{1}, Lhassane Idoumghar\textsuperscript{1}}
\IEEEauthorblockA{\textsuperscript{1}\textit{University of Haute-Alsace, IRIMAS UR 7499, F-68100 Mulhouse, France} \\
\textsuperscript{1}{\{firstname.lastname\}}@uha.fr}
\IEEEauthorblockA{\textsuperscript{2}\textit{University of Lille, CRIStAL, UMR 9189, CNRS, Centrale Lille, France} \\
\textsuperscript{2}{laetitia.jourdan}@univ-lille.fr}
}


\maketitle

\begin{abstract}
This study proposes a novel multi-objective integer programming model for a collision-free discrete drone path planning problem. Considering the possibility of bypassing obstacles or flying above them, this study aims to minimize the path length, energy consumption, and maximum path risk simultaneously. The static environment is represented as 3D grid cells. Due to the NP-hardness nature of the problem, several state-of-the-art evolutionary multi-objective optimization (EMO) algorithms with customized crossover and mutation operators are applied to find a set of non-dominated solutions.  The results show the effectiveness of applied algorithms in solving several generated test cases.
\end{abstract}


\section{Introduction}
In recent years, the demand for Unmanned Aerial Vehicles (UAVs), which is also known as drone, in both military and civilian applications is increasing rapidly due to the unavailability of human workforces to perform tedious and extensive tasks. Operating UAVs in urban areas surrounded by a large number of buildings is a tremendous challenge to assuring safety, especially for surveillance missions or delivery. Optimizing different aspects of UAV-based operations such as locating the launch stations \cite{golabi2017edge}, locating the refuelling stations \cite{shavarani2019congested}, determining the flying routes \cite{dorling2016vehicle}, and planning the flying path \cite{ghambari2018comparative} has been scrutinized in several research studies. 
\par
The path planning problem has been one of the most essential aspects of autonomous UAVs’ navigation \cite{samar2012optimal}. The main goal is to steer the UAV from the start position towards the destination via the shortest collision-free path. The path planning could be categorized into static and dynamic problems. While all the obstacles are considered to be fixed in static problems, moving obstacles transform the problem into a dynamic one \cite{bekhti2016path}. When the environment is static and known beforehand, the flight path could be well designed offline \cite{nikolos2003evolutionary}. Otherwise, the UAV needs to intelligently plan its path online \cite{peng2012intelligent}. 
\par
As a matter of fact, the UAV is supposed to reach the goal with a minimum flying distance and power consumption as well as safety maximization. Hence, a suitable path planning strategy must be designed not only to improve the effectiveness of the system but also to communicate with other elements in order to comply with the mission requirements. This purpose entails a deep analysis of various contributing techniques for implementing an effective method. 
\par
Previous studies have presented a series of techniques to tackle the aforementioned problem based on different necessities such as performance optimization, collision avoidance, real-time planning, and safety maximization. They have considered the static obstacles as the prohibited airspaces to guarantee the collision-freeness by bypassing the obstacles.  Generally, we can categorize the existing works into classical techniques (i.e., graph-based search methods, sampling-based approaches, potential field), computational intelligence (CI) methods, and hybrid approaches. Graph-based searches (e.g., A* and Dijkstra) were developed to find the shortest path between two nodes of connected graphs using a greedy logic. One of the positive characteristics of these methods is their simplicity, which implies reduced computational time. They have deterministic nature and guarantee to find the optimal collision-free path, if it exists. On the other hand, sampling-based approaches, such as Probabilistic Roadmaps (PRM) \cite{kavraki1994probabilistic} and Rapidly-exploring Random (RRT) \cite{lavalle2001randomized} have proven to be an effective framework suitable for high-dimensional spaces to produce feasible solutions; nevertheless, they do not guarantee the optimality of the solution \cite{gammell2014informed}. However, these algorithms are the most common and popular search techniques in the case of minimizing the path length as the only objective function. In recent years, CI methods including fuzzy systems, neural networks, and evolutionary algorithms (EAs) have received most of the research effort for solving the UAV path planning problem \cite{zhao2018survey}. They attract the attention of researchers because of: a) their flexibility to solve large-scale complex problems, b) their ability to apply different learning strategies to perform an effective search towards the global optimum, and c) employing for both single and multiple UAVs using different objectives.
\par
It is incontrovertible that in some cases, flying above the static obstacles results in obtaining shorter path lengths. As mentioned before, in the majority of studies in the literature, static obstacles are considered as prohibited airspace and therefore, shortening the path length by adjusting the altitude and flying above them is not possible. As a result, the majority of similar studies have considered a fixed flying altitude to bypass the obstacles. This study aims to overcome this shortcoming by considering both possibilities of bypassing the altitudes or flying above them. To do so, a new multi-objective mathematical formulation that accounts for both mentioned possibilities is proposed to minimize the path length, energy consumption and path risk simultaneously. To calculate the energy required to travel from point \(\textit{i}\) to point \(\textit{j}\) at different altitude levels, the formulation proposed by \cite{dorling2016vehicle} is used. To account for the required energy for increasing the altitude, the term (\(\textit{W}{g}{\Delta}_{ij}^{+}\)) is added to the mentioned formulation:

\begin{equation}\label{eq1}
\textstyle \theta = W^{3⁄2} \sqrt{\frac{g^{3}}{2\rho\zeta n}}{ }     \frac{{d}_{ij}}{\nu} + {W}{g}{\Delta}_{ij}^{+}
 \end{equation}
\par
 Where \(\textit{W}\) is the drone and battery weight (\(\textit{kg}\)), \(\textit{g}\) is the gravity (\(\textit{N}\)), \({d}_{ij}\) is the distance between point \(\textit{i}\) to point \(\textit{j}\) (\(\textit{m}\)), \(\textit{}{\Delta}_{ij}^{+}\) is the increased altitude (\(\textit{m}\)), \(\textit{}{\nu}\) is the flying speed (\(\textit{m/s}\)), \(\zeta\) is the area of spinning blade disc in \(\textit{m}^2\), and \(\textit{n}\) is the number of rotors. 
\par 
The effect of altitude level on the gravity could be neglected as the gravitational force would have decreased just by 1.2\% if an object flies at an altitude of 40 km. Since the flying altitude of drones is considerably lower that what mentioned here, the effect of altitude on the gravity could be ignored. The other factor that changes by altitude is fluid density of air. As described by \cite{paredes2017study}, the fluid density of air in different altitudes could be calculated according to Eq. \ref{eq2} in which \(\textit{H}\) is the altitude in \(\textit{meters}\).

\begin{equation}\label{eq2}
\textstyle \rho = (1-2.2558.10^{-5}H)^{4.2577}
 \end{equation}

Furthermore, this study proposes a new solution representation, along with customized search operators in order to solve the problem in a flexible manner. To do so, several evolutionary multi-objective optimization algorithms including Non-dominated Sorted Genetic Algorithm (NSGA-II), Reference-point based Non-dominated Sorting Genetic Algorithm (NSGA-III), and Strength Pareto Evolutionary Algorithm (SPEA-II) are applied on five generated test cases. The obtained results demonstrate how new modeling with customized operations can efficiently help the algorithms to find the Pareto-regions, especially in cluttered spaces.  
\par
The rest of this paper is organized as follows. The problem definition is given in section \ref{problem def}. Section \ref{method} starts with basic concepts of several multi-objective algorithms which are used as base line of this empirical study in subsection \ref{AlgConcepts}. Next, in subsections \ref{IMPD} the details of implementation including environment modeling, solution representation, and search operators (initialization, crossover, and mutation) are provided. The simulation results and discussion are presented in Section \ref{experiments}. Finally, the paper is concluded in Section \ref{Conc}.

\section{Problem definition}\label{problem def}
The aim of this study is to find the drone’s best path while traveling from point A to point N in an environment that is replete with different obstacles of different heights such that the total traveling distance, energy consumed and the maximum path risk are minimized simultaneously. In order to have a comprehensive knowledge of the environment and identify the obstacles, the ground surface is decomposed into several small cells and the flying space is decomposed into several altitude levels. The cells containing the obstacles and the concomitant altitude level of the obstacle are identified. The drone starts the mission from point A and aims to reach point N in a manner that the mentioned objective functions are minimized. For each cell of the ground surface, the set of succeeding cells could be determined by allowing the drone to fly in five directions: to the north, to the south, to the east, to the north-east, and to the south-east. The distance between each cell and its succeeding cells is calculated in advance. While arriving at each cell, if there exists a succeeding cell containing an obstacle, the drone has the options of choosing other succeeding neighbour cells to take a detour and avoid passing through the obstacle, or setting the altitude according to the height of the obstacle and pass above it. It is obvious that if all the succeeding cells contain obstacles, the drone must adjust the altitude and pass above one of the neighbour cells. It is noteworthy to mention that the adjusted altitude while passing the cells could also be restricted. While passing across some areas, the authorization may restrict the altitude due to different reasons such as the proximity to military bases, airports, or traditional flying corridors. To find better solutions in terms of the mentioned objective functions, in this study the drone is allowed to decrease its altitude from the previously adjusted levels, if applicable. The drone’s energy consumption depends on the weight, travelled distance, and the fluid density of air. Since the fluid density of air changes with altitude, this study accounts for the real energy consumption based on both flying distance and flying altitude. For each two consecutive visited cells, the average altitude is used to calculate the fluid density of air. Additionally, in order to reflect a more realistic image of energy consumption, this study considers the required energy to increase the altitude while passing the obstacles. The direct distance between the cells is calculated from the north-west corner of the cells. The altitude-change procedure (if applicable) starts from the north-west corner of the current cell towards the north-west corner of the selected succeeding cell. So, before arriving at any cell, the altitude of the drone has already been adjusted. Therefore, considering the change in the altitude and the direct distance, the real travel distance between the cells is calculated using the Pythagoras theorem. It is assumed that the drone’s flying speed is fixed for the entire route. Based on the data of the other organizations’ drone routes and the recorded air traffic of the environment, a risk factor is assigned to all the consecutive cells that account for the collision probability at different altitude levels. For each visiting cell, the maximum risk factor of traversed altitude levels while passing to the successive cell, is calculated. The accumulated maximum path risk is considered as the third objective function.   

\section{Mathematical formulation}\label{mathform}
In order to develop the mathematical model, the following sets, scalars, parameters, and variables are used:\\

Sets:\\
\($\textit{I}$\): the set of ground cells (\($\textit{i}$\), \($\textit{j}$\), \($\textit{g}$\), A (the starting cell), \($\textit{N}$\) (the final destination) \(\in\) \($\textit{I}$\))\\
\($\textit{K}$\): the set of altitude levels (\($\textit{k}$\) ,\($\textit{k'}$\) $\in$ \($\textit{K}$\)) \\
\({\delta}_{i}^+\): the sets of cell (\($\textit{i}$\)’s succeeding cells) \\
\({\delta}_{i}^-\): the set of cell (\($\textit{i}$\)’s preceding cells) \\

Scalars:\\
\(\textit{M}\) : a large positive number\\
\(\textit{v}\): the drone’s flying speed\\
\(\theta\): a scalar used in calculation of consumed energy \( (\theta = W^{3⁄2} \sqrt{g^{3}/2\zeta n}) \)  \\


Parameters:\\
\({a}_{j}\): the length of obstacle located in cell \(\textit{j}\), if exists\\
\({U}_{j}\): the maximum allowed flying altitude over cell \(\textit{j}\)\\
\({r}_{ij}^k\): the risk factor concomitant to flying from cell \(\textit{i}\) to cell \(\textit{j}\) at altitude \(\textit{k}\)\\
\({h}^A\): the drone’s starting altitude\\
\({h}^k\): the altitude of level \(\textit{k}\)\\
\({\rho}^k\): the fluid density of air while flying at altitude level \(\textit{k}\)\\
\({\rho}^A\): the fluid density of air while starting flying at the first cell\\
\({a}_{ij}\): the direct distance between the neighbour cells \(\textit{i}\) and \(\textit{j}\)\\

Decision variables:\\
\({X}_{ij}^k\): $1$ if drone enters cell \(\textit{j}\) from cell \(\textit{i}\) at an altitude of level \(\textit{k}\); $0$ otherwise\\
\({\Delta}_{ij}\): the real altitude change while flying from cell \(\textit{i}\) to cell \(\textit{j}\) (\({\Delta}_{ij}\)  free in sign)\\
\({\Delta}_{ij}^+\): the height ascended while flying from cell \(\textit{i}\) to cell \(\textit{j}\) (\({\Delta}_{ij}^+\)  \(\geq\)  $0$)\\
\({\Delta}_{ij}^-\): the height descended while flying from cell \(\textit{i}\) to cell \(\textit{j}\) (\({\Delta}_{ij}^-\)  \(\geq\)  $0$)\\

The first set of constraints are called the network flow constraints:

\begin{equation}\label{eq3}
\textstyle \sum_{j\in {\delta}_{A}^+} \sum_{k\in K} {X}_{ij}^k = 1
\end{equation}

\begin{equation}\label{eq4}
\textstyle \sum_{j\in {\delta}_{N}^-} \sum_{k\in K} {X}_{iN}^k = 1
\end{equation}

\begin{equation} \label{eq5}
\begin{split}
\textstyle \sum_{j\in {\delta}_{i}^+} \sum_{k\in K} {X}_{ij}^k & = \textstyle \sum_{j\in {\delta}_{i}^-} \sum_{k\in K} {X}_{ji}^k  \\
& \forall i\in I \arrowvert \{i\neq A \quad and \quad i\neq N\}
\end{split}
\end{equation}

\begin{equation}\label{eq6}
\textstyle \sum_{j\in {\delta}_{N}^+} \sum_{k\in K} {X}_{Nj}^k = 0
\end{equation}

Eq. \ref{eq3} and Eq. \ref{eq4} indicate that the drone must leave the starting cell and arrive at the final destination. Eq. \ref{eq5} assures that except for the starting cell and final destination, the drone should leave the entered cells. Eq. \ref{eq6} guarantees that after arriving at the final destination the path is finished.
The following set of constraints accounts for the designated altitude level while arriving at each cell. Using the designated altitude level, the change in altitude for all visited pairs of cells could be calculated. 

\begin{multline} \label{eq7}
\textstyle \sum_{k\in K} {h}^k {X}_{ij}^k \geqslant {a}_{j} - M(1-\sum_{k\in K} {X}_{ij}^k) \\
\forall j\in I  \arrowvert \{j\neq A\},\forall i\in {\delta}_{j}^-
\end{multline}

\begin{equation} \label{eq8}
\begin{split}
\textstyle \sum_{k\in K} {h}^k {X}_{ij}^k \leqslant {U}_{j} \quad \quad
\forall j\in I  \arrowvert \{j\neq A\},\forall i\in {\delta}_{j}^-
\end{split}
\end{equation}

Eq. \ref{eq7} assures that in the case of traveling from cell \(\textit{i}\) to cell \(\textit{j}\),  the adjusted altitude level while arriving at cell \(\textit{j}\) is higher than the located obstacle (if any). Eq. \ref{eq8} defines an upper bound for the adjusted altitude level while arriving at cell \(\textit{j}\).

The change in altitude while traveling from cell \(\textit{i}\) to cell \(\textit{j}\) is calculated as follows:

\begin{equation} \label{eq9}
\begin{split}
{\Delta}_{Aj} = \textstyle \sum_{k\in K} ({h}^k-{h}^A) {X}_{Aj}^k \quad\quad \forall j\in {\delta}_{A}^+
\end{split}
\end{equation}

\begin{equation} \label{eq10}
\begin{split}
{\Delta}_{ij} = \textstyle \sum_{g\in {\delta}_{i}^-} \sum_{k\in K} \sum_{k'\in K} ({h}^k-{h}^{k'}) {X}_{ij}^k {X}_{gi}^{k'} \\
\forall i\in I \arrowvert \{i\neq A\},\forall j\in {\delta}_{i}^+, \forall g\in {\delta}_{i}^-
\end{split}
\end{equation}

Due to the multiplication of decision variables, Eq. \ref{eq10} is non-linear. In order to linearize this formulation, assuming \( {X}_{ij}^k {X}_{gi}^{k'} = {U}_{gij}^{kk'}\), Eq. \ref{eq10}  is replaced by the following constraints:

\begin{equation} \label{eq11}
\begin{split}
{\Delta}_{ij} = \textstyle \sum_{g\in {\delta}_{i}^-} \sum_{k\in K} \sum_{k'\in K} ({h}^k-{h}^{k'}) {U}_{gij}^{kk'} \\
\forall i\in I \arrowvert \{i\neq A\},\forall j\in {\delta}_{i}^+, \forall g\in {\delta}_{i}^- 
\end{split}
\end{equation}

\begin{equation} \label{eq12}
\begin{split}
{U}_{gij}^{kk'} \geqslant {X}_{ij}^k + {X}_{gi}^{k'} -1\\
\forall i\in I \arrowvert \{i\neq A\},\forall j\in {\delta}_{i}^+, \forall g\in {\delta}_{i}^-
\end{split}
\end{equation}

\begin{equation} \label{eq13}
\begin{split}
2{U}_{gij}^{kk'} \leqslant {X}_{ij}^k + {X}_{gi}^{k'}\\
\forall i\in I \arrowvert \{i\neq A\},\forall j\in {\delta}_{i}^+, \forall g\in {\delta}_{i}^-
\end{split}
\end{equation}

Using the real change in altitude obtained from Eq. \ref{eq11} to Eq. \ref{eq13}, the ascended and descended altitude while traveling from cell \(\textit{i}\) to cell \(\textit{j}\) is calculated according to the following set of constraints in which \({y}_{ij}\) and \({y}_{ij}^{'}\) are the binary variables:

\begin{equation} \label{eq14}
\begin{split}
{y}_{ij}{\Delta}_{ij}^{+} -{y}_{ij}^{'}{\Delta}_{ij}^{-}  = {\Delta}_{ij} \quad\quad \forall i\in I,\forall j\in {\delta}_{i}^+
\end{split}
\end{equation}

\begin{equation} \label{eq15}
\begin{split}
{\Delta}_{ij}^{+} - {\Delta}_{ij}^{-}  = {\Delta}_{ij} \quad\quad \forall i\in I,\forall j\in {\delta}_{i}^+
\end{split}
\end{equation}

\begin{equation} \label{eq16}
\begin{split}
{y}_{ij} + {y}_{ij}^{'} = 1 \quad\quad \forall i\in I,\forall j\in {\delta}_{i}^+
\end{split}
\end{equation}

Again, due to the multiplication of decision variables, Eq. \ref{eq14} is non-linear. Therefore, considering \({\wp}_{ij}^{+}={y}_{ij} {\Delta}_{ij}^{+}\) and \({\wp}_{ij}^{-}={y}_{ij}^{'}{\Delta}_{ij}^{-}\), Eq. \ref{eq14} is linearized by replacing to the following set of constraints:

\begin{align} \label{eq17}
{\wp}_{ij}^{+} - {\wp}_{ij}^{-} = {\Delta}_{ij} \quad
\forall i\in I,\forall j\in {\delta}_{i}^+
\end{align}

\begin{align} \label{eq18}
{\wp}_{ij}^{+} \geqslant {\Delta}_{ij}^{+} - M(1-{y}_{ij}) \quad
\forall i\in I,\forall j\in {\delta}_{i}^+
\end{align}

\begin{align} \label{eq19}
{\wp}_{ij}^{+} \leqslant {\Delta}_{ij}^{+} + M(1-{y}_{ij}) \quad
\forall i\in I,\forall j\in {\delta}_{i}^+
\end{align}

\begin{align} \label{eq20}
{\wp}_{ij}^{+} \leqslant M{y}_{ij} \quad
\forall i\in I,\forall j\in {\delta}_{i}^+
\end{align}

\begin{align} \label{eq21}
{\wp}_{ij}^{-} \geqslant {\Delta}_{ij}^{-} - M(1-{y}_{ij}^{'}) \quad
\forall i\in I,\forall j\in {\delta}_{i}^+
\end{align}

\begin{align} \label{eq22}
{\wp}_{ij}^{-} \leqslant {\Delta}_{ij}^{-} + M(1-{y}_{ij}^{'}) \quad
\forall i\in I,\forall j\in {\delta}_{i}^+
\end{align}

\begin{align}\label{eq23}
{\wp}_{ij}^{+} \geqslant M{y}_{ij}^{'} \quad
\forall i\in I,\forall j\in {\delta}_{i}^+
\end{align}


The first objective function is to minimize the path length. Considering a triangle in which the base is the direct distance between the visited consecutive cells and the height is the absolute value of altitude change, the drone’s travelled distance is calculated using the Pythagoras theorem. So, the first objective function that minimizes the total travelled distance is:


\begin{multline}\label{eq24}
Min \quad {Z}_{1} = \textstyle \sum_{j\in {\delta}_{A}^+} \sum_{{k}\in {K}} (({h}^{k}-{h}^A)^{2}+{d}_{Aj}^{2})^{1/2}{X}_{Aj}^{k} \\ +
\textstyle \sum_{{i}\in {I} \arrowvert {i}\neq {A}} \sum_{{j}\in {\delta}_{i}^+} \sum_{{g}\in {\delta}_{i}^-} \sum_{{k}\in {K}} \sum_{{k'}\in {K}}\\
(({h}^{k}-{h}^{k'})^{2}+{d}_{ij}^{2})^{\frac{1}{2}}{U}_{gij}^{kk'}
\end{multline}

The energy consumption while traveling from cell \(\textit{i}\) to cell \(\textit{j}\) depends on the related fluid density of air and the travelled distance. The fluid density of air is a function of flying altitude. In this research, the fluid density between cells \(\textit{i}\) and \(\textit{j}\) is considered as the average of fluid density of air while arriving at cells \(\textit{i}\) and \(\textit{j}\). So, the second objective function could be written as:


\begin{multline}\label{eq25}
Min \quad {Z}_{2} = \textstyle \sum_{{j} \in {\delta}_{A}^+} \sum_{{k}\in {K}} \frac{\theta}{\nu}(\frac{({h}^{k}-{h}^{k'})^{2}+{d}_{Aj}^{2}}{\frac{\rho^{A}+\rho^{k}}{2}}) ^{1/2}{X}_{Aj}^{k} \\
+ \textstyle \sum_{i\in I \arrowvert i\neq A}\sum_{j\in {\delta}_{i}^+}\sum_{g\in {\delta}_{i}^-} \sum_{{k}\in {K}} \sum_{{k'}\in {K}} \frac{\theta}{\nu} \\
(\frac{({h}^{k}-{h}^{k'})^{2}+{d}_{ij}^{2}}{\frac{\rho^{k'}+\rho^{k}}{2}})^{\frac{1}{2}}{U}_{gij}^{k}{k'}+ \textstyle \sum_{{i} \in {I}} \sum_{{j} \in {\delta}_{i}^+} {Wg} {\Delta}_{ij}^{+}
\end{multline}

Finally, the third objective function that considers the maximum accumulated path risk is calculated as:

\begin{multline} \label{eq26}
        Min \quad {Z}_{3} = \textstyle \sum_{j\in {\delta}_{A}^+} \sum_{k\in K} \displaystyle \max_{{\Phi \in [(\min(k,A)),(\max(k,A))]}}\\
        \{{r}_{A}^{\Phi}\} {X}_{Aj}^{k} 
        +\textstyle \sum_{i\in I \arrowvert i\neq A} \sum_{j\in {\delta}_{i}^+} 
        \sum_{g\in {\delta}_{i}^-} \sum_{k\in K} \sum_{k'\in K}\\
        (\max_{{\Phi \in [(\min(k,k')),(\max(k,k'))]}} \{{r}_{i}^{\Phi}\} {U}_{gij}^{kk'})
\end{multline}


\section{Methodology}\label{method}
The path planning problem for autonomous mobile robots is NP-hard \cite{li2012efficient}. Therefore, several intelligent optimization methods have been successfully used for solving this problem, especially while dealing with more than one objective function. This section explains how the proposed problem can be solved using evolutionary multi-objective optimization algorithms. For this purpose, several common algorithms including NSGA-II,  NSGA-III, and SPEA-II are selected. These algorithms are similar in search operations. First, these algorithms are briefly described. Thereafter, specific operations are elaborated in more detail.

\subsection{Preliminary concepts of algorithms}\label{AlgConcepts}
The strength pareto evolutionary algorithm (SPEA) is proposed by Zitzler et al. \cite{zitzler1998evolutionary} in order to approximate the non-dominated solutions for multi-objective optimization problems. SPEA-II \cite{zitzler2001spea2} is an improved elitist version of SPEA, which incorporates in contrast to its predecessor a fine-grained fitness assignment strategy, and techniques for both archive truncation and density-based selection.

\par
NSGA-II is one of the most popular multi-objective optimization algorithms presented by Deb et al. \cite{deb2000fast} which uses an elitist principle. It has three special characteristics, fast non-dominated sorting approach, an explicit diversity preserving mechanism (crowded distance estimation procedure) and simple crowded comparison operator to emphasize the non-dominated solutions. It is claimed that this technique outperformed SPEA in terms of finding a diverse set of solutions.
\par

Deb and Jain \cite{deb2013evolutionary} developed the well-know NSGA-III in early 2014, to deal with both unconstrained and constrained many objective optimization problems. It is based on an external guidance mechanism to maintain diversity among its solutions. The main difference between the two NSGA-II and NSGA-III is that NSGA-III employed a set of reference points to keep the exploration of the Pareto points during the search. It yields better results, even distribution of Pareto points across the objective space, even when the number of objectives is large.

\subsection{Implementation details}\label{IMPD}
\par
The developed algorithms initiate with a randomly generated population of solutions of size \(\textit{N}\). Each solution, which is called a chromosome, has a variable length and could be represented as a matrix of two rows. The first-row alleles determine the index of visiting cells while the corresponding second-row alleles indicate the adjusted altitude just before entering that cell. To generate the first-row alleles, for each cell of the ground space, a set of neighbourhood cells that directs the mentioned cell to the destination is generated. The first-row starts with the index of the origin. Then, the next gene is filled with a randomly selected cell index in the neighbourhood set of the origin. Up to reaching the destination, the next genes are filled by the not-used randomly selected neighbour cells of the previous gene. If all the neighbour cells are used to fill the previous genes, the destination is not reached and the solution will be infeasible. For each first-row gene, the corresponding second-row gene is filled by a randomly selected altitude level that is in the range of the selected cell’s obstacle height (if any) and its maximum allowed flying altitude. To generate the reasonable altitude levels, adjusting the altitude to the height of entering cell’s obstacle height (if any), keeping the previous cell’s altitude level (if it is greater than the height of the obstacle existed on the entering cell, if any), or selecting an altitude level in the mentioned range in a random manner have the same probabilities of being applied.
\par
\begin{figure}[htp]
    \centering
    \includegraphics[width=6.5cm]{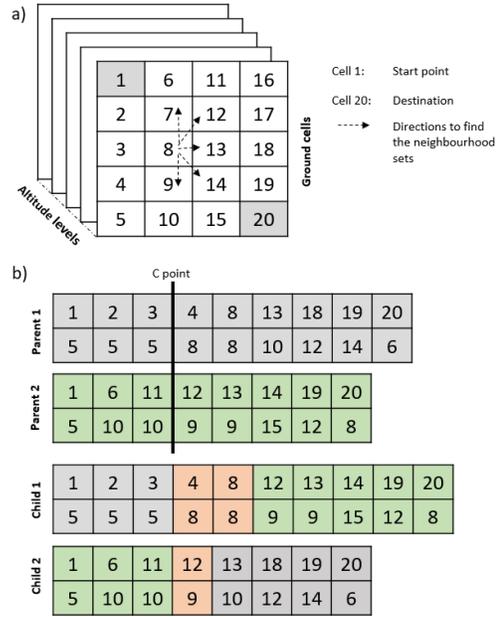}
    \caption{a) The flying environment; b) The crossover operation}
    \label{fig:Fig1.pdf}
\end{figure}

The one-point crossover operation applied in this study reproduces two offspring chromosomes by mating two randomly selected parental chromosomes. The crossover point is selected randomly along the length of the mating chromosomes and decomposes them into two segments. Suppose that C is the position of the crossover point. Considering the first-row genes, if the \(\textit{C+1}^{th}\) allele of the second parent is among the neighbourhood set of the \(\textit{C}^{th}\) allele of the first parent, the first child is generated by merging all the first segment of the first parent with all the second segment of the second parent. Otherwise, up to reaching a point by which merging the parents is possible, crossover point will be iteratively shifted to the right-hand side for the selected parent (see child 1 in Fig. \ref{fig:Fig1.pdf}.b) or for both of the parents (see child 2 in Fig. \ref{fig:Fig1.pdf}.b)
\par
Since any further change in the first-row genes requires a corresponding search and repairing procedure that was explained in crossover section, the mutation operation is applied to the second-row genes. Based on the mutation probability, a random number of second-row genes are selected and using the probabilistic procedure previously mentioned about filing the second-row genes, their alleles are replaced with randomly generated numbers.

\begin{figure}
    \centerline{\includegraphics[width=0.48\textwidth]{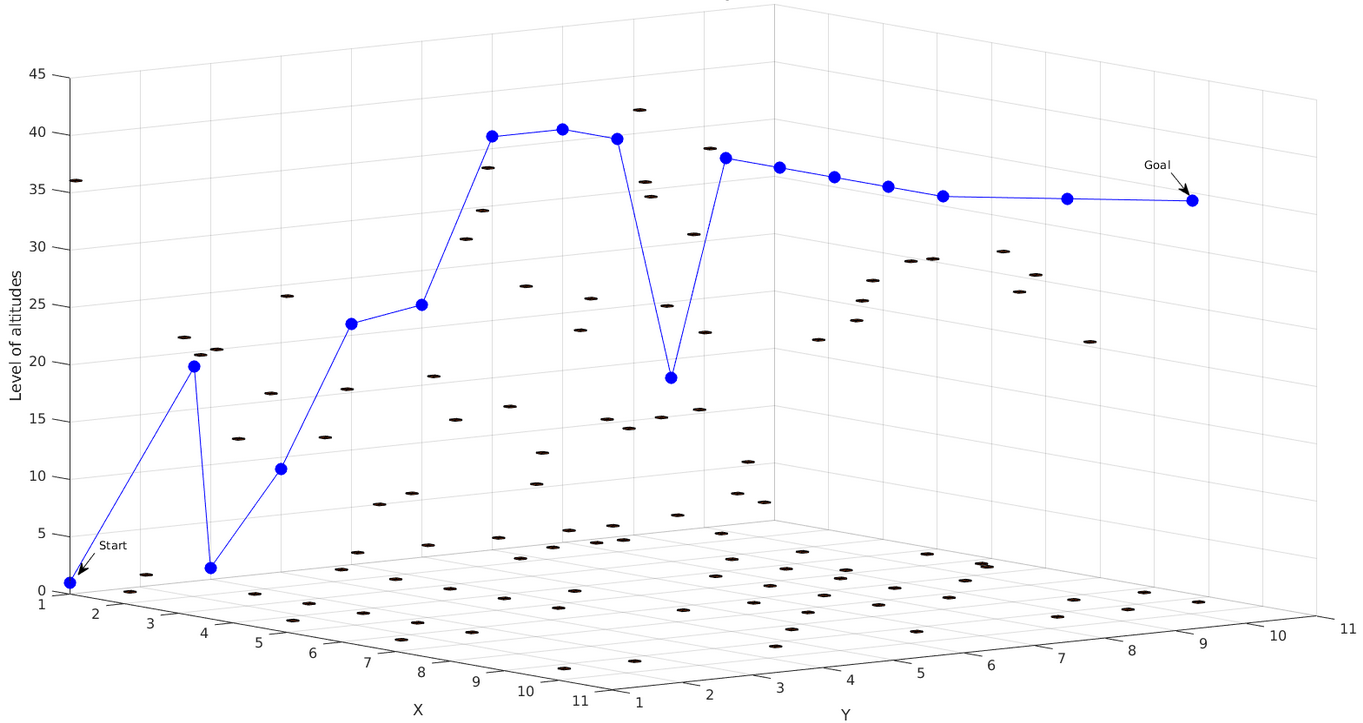}}
    \caption{The obtained path from pareto-front set with NSGA-II}
    \label{fig:Fig3.PNG}
\end{figure}

To show the utility of the modeling and customized operators, a simple 10*10 environment is considered and the concomitant path obtained from one of the members of the best Pareto-front of NSGA-II is illustrated in Fig. \ref{fig:Fig3.PNG}. The dashes show the altitude of each cell's obstacle. As shown in this figure, in some cases, the drone has bypassed the obstacles by selecting alternative cells, while in some other cases, the drone has flown above the obstacles by adjusting the altitude.

\section{Experiment results}\label{experiments}

\subsection{Experimental protocol}
The simulations are executed to analyze and prove the utility and validity of the presented mathematical modeling in this study. To do so, several EMO algorithms including NSGA-II, NSGA-III, and SPEA-II are applied on five randomly generated 3D test cases in order to accomplish a series of experiments. There are five test cases which are different in environment size. Each test case has four different versions that differ in the number of obstacles, starting point, and destination. The obstacles are randomly distributed in the area. 
Test cases are available in the GitHub library\footnote{({https://github.com/Sghambari/3D-dataset-for-mathematical-modelling.git})}.

\par
The applied algorithm parameters are the population size, crossover probability ($CR$), mutation probability ($MP$), and mutation rate ($mu$) with their permissible range values. Since the parameters influence the efficiency and effectiveness of applied algorithms, it is necessary to adjust them in advance to implement the algorithms. Hence, instead of using a trial-and-error approach to identify good values for them, \textit{MAC} method~\cite{rakhshani2019mac} is utilized as an automated algorithm configuration tool for multi-objective methods in order to obtain the best parameter values. In \textit{MAC}, the maximum budget of 100 function evaluations or reaching 48 hours of computation is used for each run. 

\par

This section first investigates the correlation between path length and energy consumption objective functions. Indeed, it is explained that the weighted sum approach is utilized to combine these objective functions. Thereafter, the results of applying the tuned and non-tuned algorithms on the generated test cases are reported.

\subsection{weighted sum approach for path length and energy consumption objectives}

\begin{figure*}[h]
\begin{subfigure}{.5\textwidth}
  \centering
  \includegraphics[width=\linewidth]{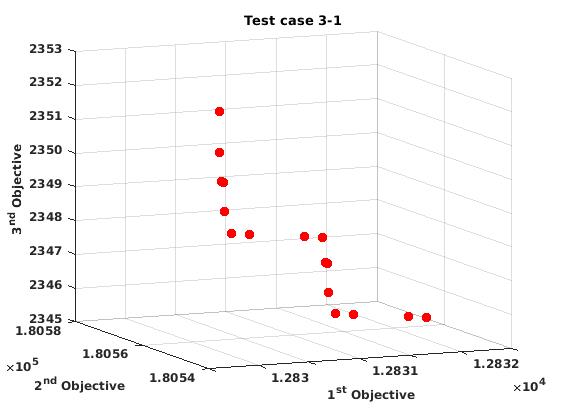}
  \caption{}
\end{subfigure}%
\begin{subfigure}{.45\textwidth}
  \centering
  \includegraphics[width=\linewidth]{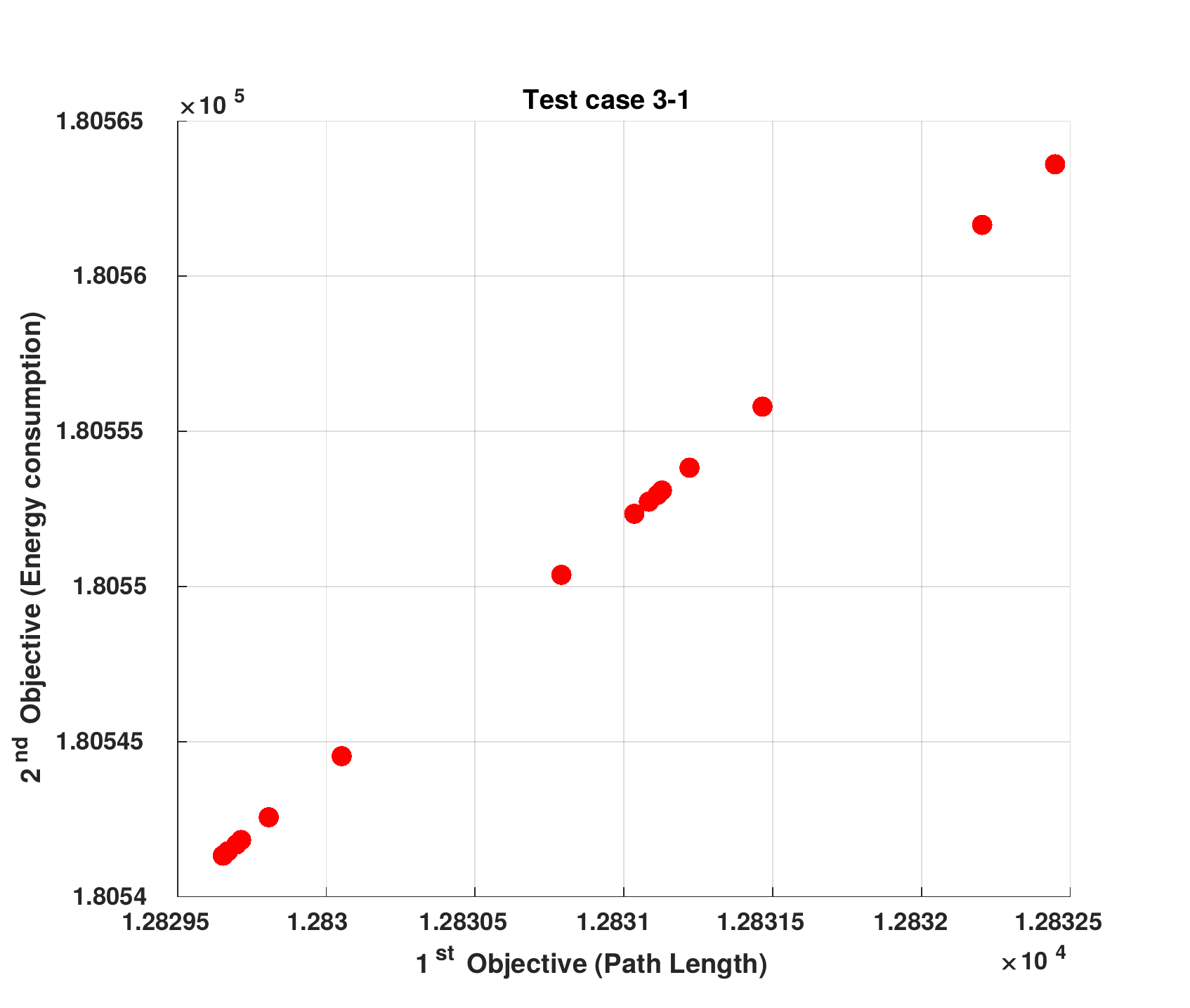}
  \caption{}
\end{subfigure}
\caption{The correlation between path length and energy consumption; the 3-D view of three objectives is shown in left-side figure and the right-side figure is for a 2-D view of path length and energy consumption}
\label{fig:Figscorrelation}
\end{figure*}

The experimental results obtained from solving different test cases using the EMO algorithms show that there is a correlation between path length and energy consumption. This correlation is illustrated in figure \ref{fig:Figscorrelation}, as an examples of one test case from both 2D and 3D view.

\par
The weighted sum approach is one of the classical methods for solving multiple non-conflicting objective functions. As it is formulated in Eq. (\ref{eq27}), this approach assigns a weight \(\textit{w}_{i}\) to each objective function \(\textit{z}_{i}(x)\) and converts the problem to a single objective one.
\begin{equation}\label{eq27}
Min Z = {w}_{1}{z}_{1}(x) + {w}_{1}{z}_{1}(x) + ... + {w}_{n}{z}_{n}(x) 
\end{equation}


where \(\sum {w}_{i} = 1\). The main challenge of applying this method is the selection of a weight vector for each run. To address this issue, Hajela and Lin \cite{hajela1992genetic} proposed a multi-criterion optimal design for multi-objective optimization in order to automate this procedure. Inspired by the mentioned design, in this study, each solution \(\textit{x}_{i}\) in the population utilizes a different weight vector in the calculation of the new objective function which is the combination of path length and energy consumption. The
weights vector of \(\textit{w}_{i}\) is embedded within the chromosome of
solution \(\textit{x}_{i}\). Therefore, multiple solutions can be simultaneously searched in a single run. Furthermore, weights
vector can be adjusted to promote the diversity of the population. 

\subsection{Results and discussions}
In this section, the computational results of implementing the EMO algorithms on the generated test cases are provided, followed by a general discussion of the utility of the mathematical modeling and the effect of automatic parameter configuration on non-dominated (Pareto sets of) solutions. There are various indicators in the literature to evaluate the quality of Pareto sets of solutions. In this study, unary hypervolume (\textit{HV}) \cite{zitzler1999multiobjective}, as one of the broadly used performance indicator for EMO algorithms, is considered as a baseline of the comparison for two-objective formulation (combination of path length and energy consumption with path risk objective).
\par
Table \ref{Tab:table1} reports the percentage of improvement of \textit{HV} for different algorithms with and without automatic parameter tuning. The default values of parameters for the selected algorithms are taken from their original papers \cite{deb2000fast,deb2013evolutionary,zitzler2001spea2}. As can be seen, the best results are obtained by employing \textit{MAC}. Among the compared algorithms with \textit{MAC}, for test case 1, SPEA-II outperforms the other competitors. For test cases 2 and 3, NSGA-II and NSGA-III have the same performance. However, NSGA-III had a better performance for test case 4. For test case 5, NSGA-II could win with three 100\% of improvement in comparison to NSGA-III and SPEA-II. Also, for the reported results without parameter tuning, SPEA-II outperforms in test cases 1 and 2. For test case 3 NSGA-III and for test case 4, NSGA-II had better performances. Finally, for test case 5, both NSGA-II and NSGA-III have the same performance. Generally, it can be concluded that the fine-tuned NSGA-II algorithm outperforms the other competitors in most of the test cases.
\par
\begin{figure*}[h]
\begin{subfigure}{.35\textwidth}
  \centering
  \includegraphics[width=\linewidth]{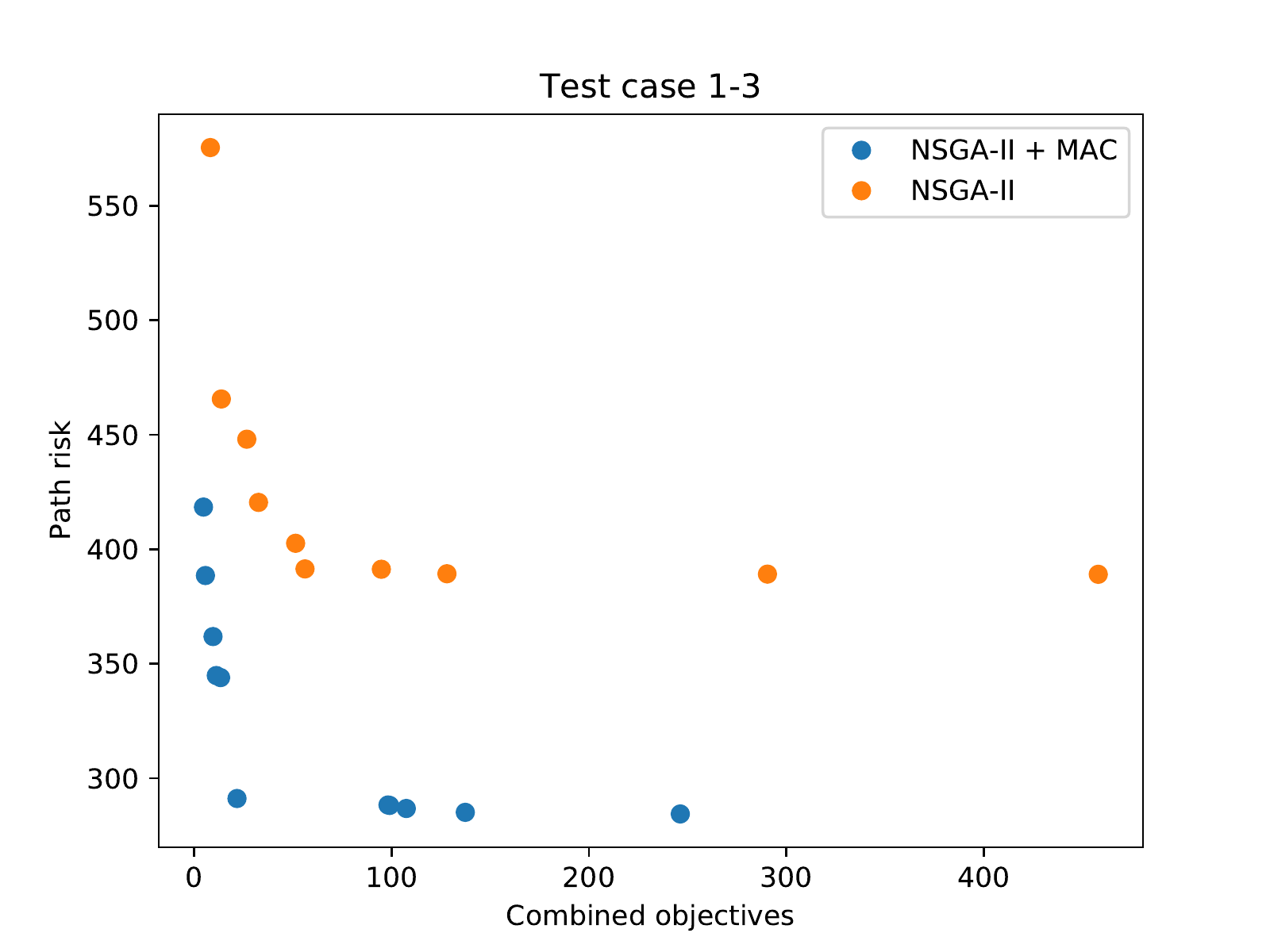}
  \caption{}
\end{subfigure}%
\begin{subfigure}{.35\textwidth}
  \centering
  \includegraphics[width=\linewidth]{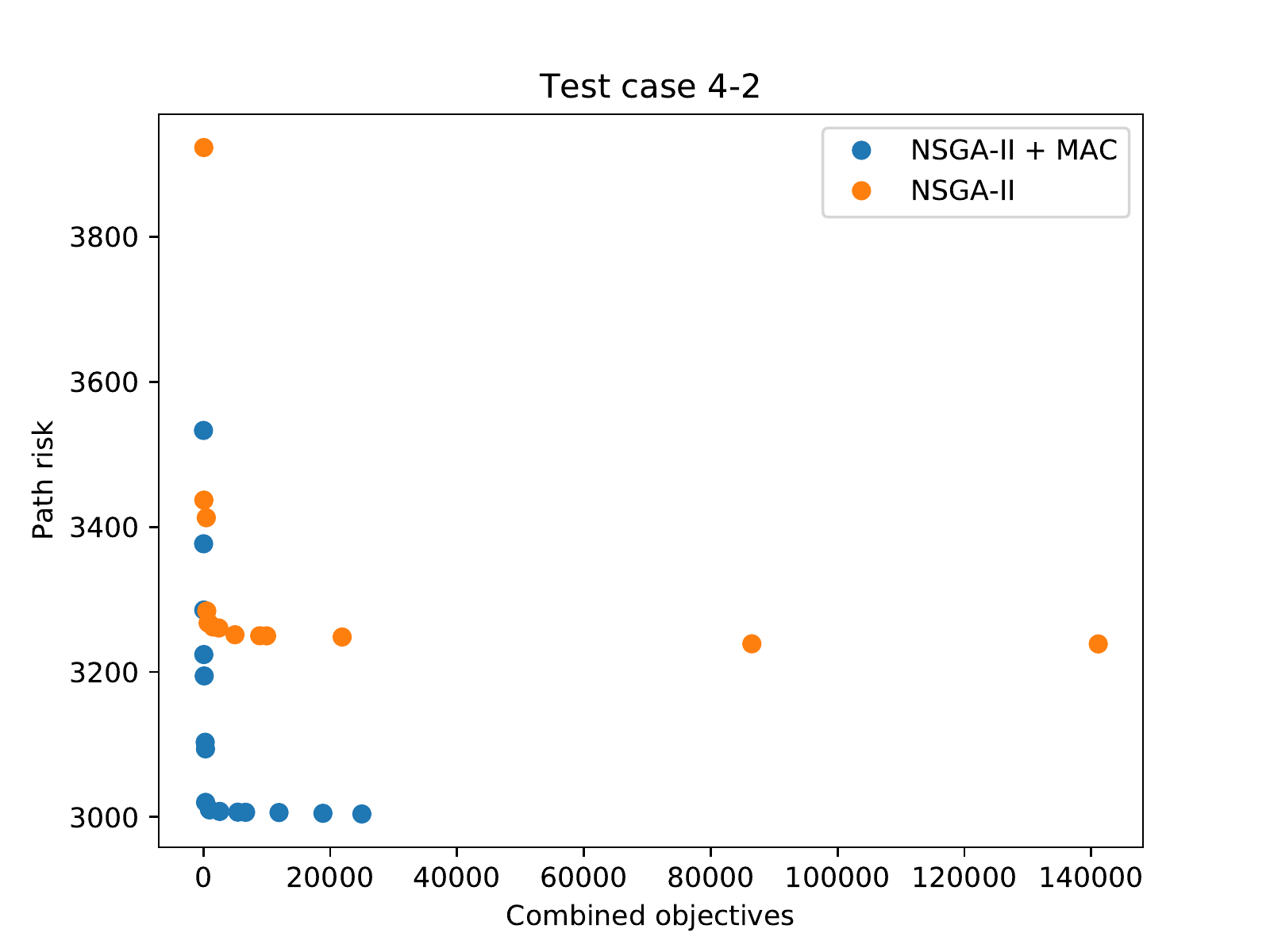}
  \caption{}
\end{subfigure}
\begin{subfigure}{.35\textwidth}
  \centering
  \includegraphics[width=\linewidth]{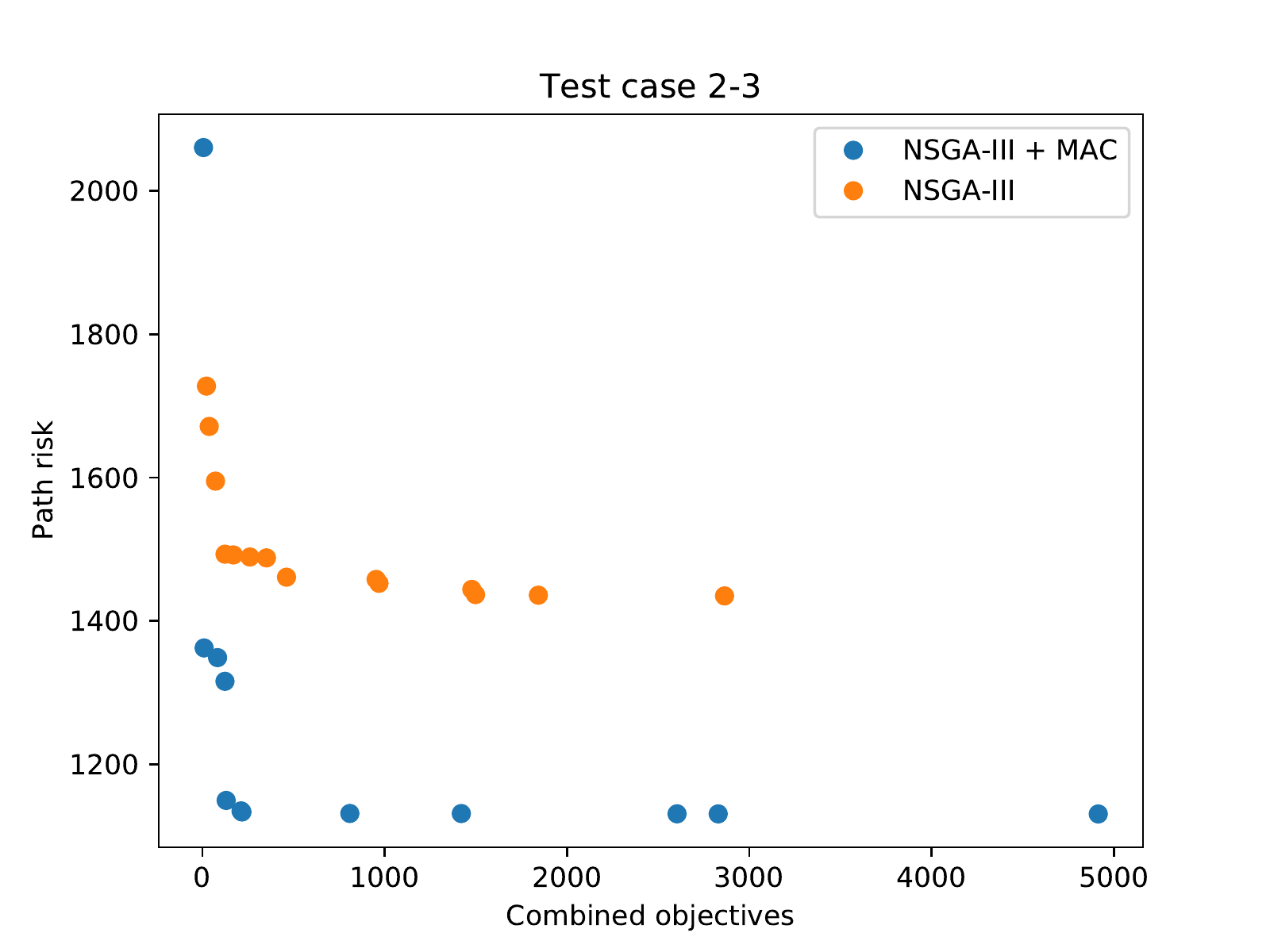}
  \caption{}
\end{subfigure}%

\begin{subfigure}{.35\textwidth}
  \centering
  \includegraphics[width=\linewidth]{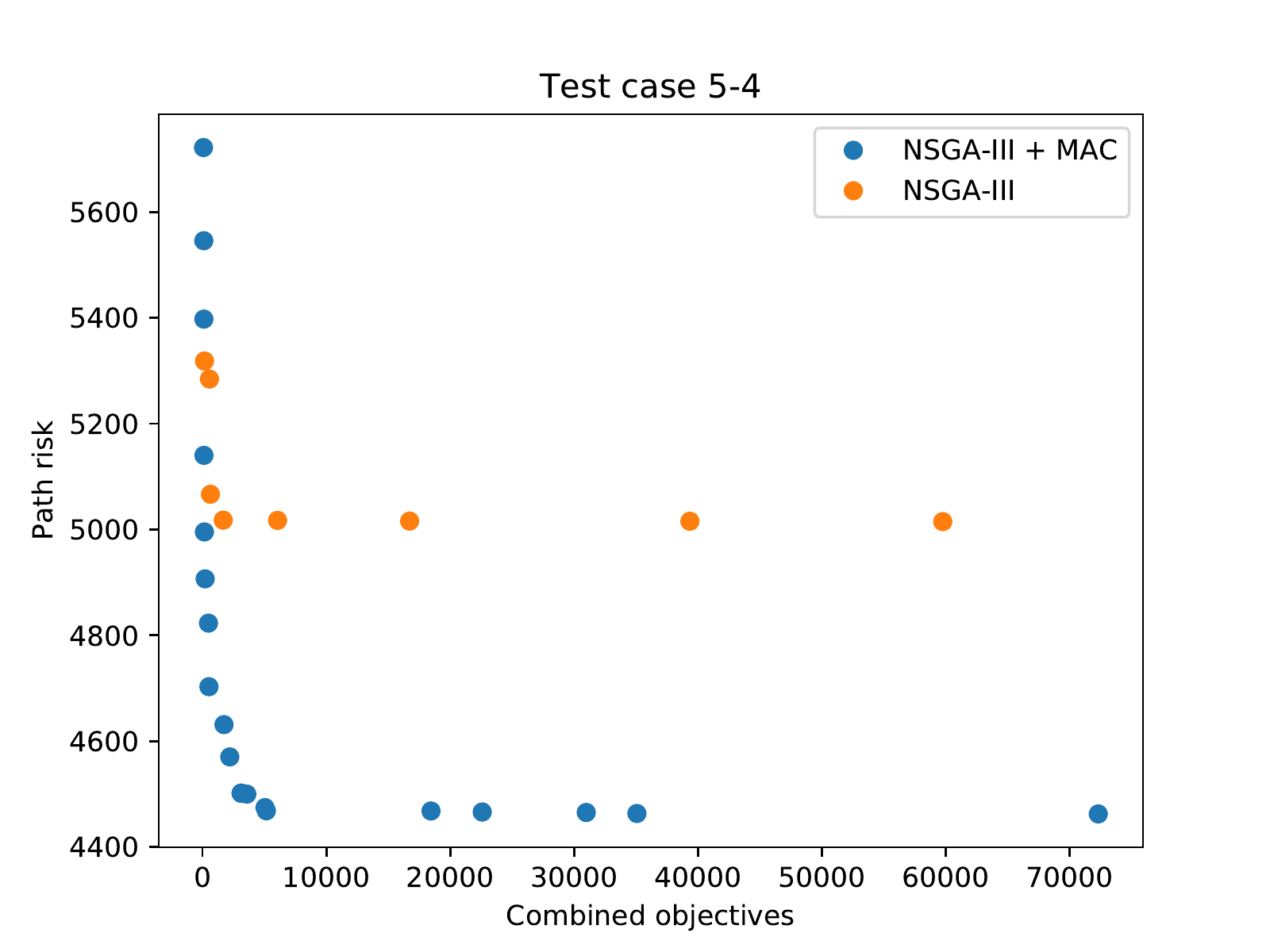}
  \caption{}
\end{subfigure}
\begin{subfigure}{.35\textwidth}
  \centering
  \includegraphics[width=\linewidth]{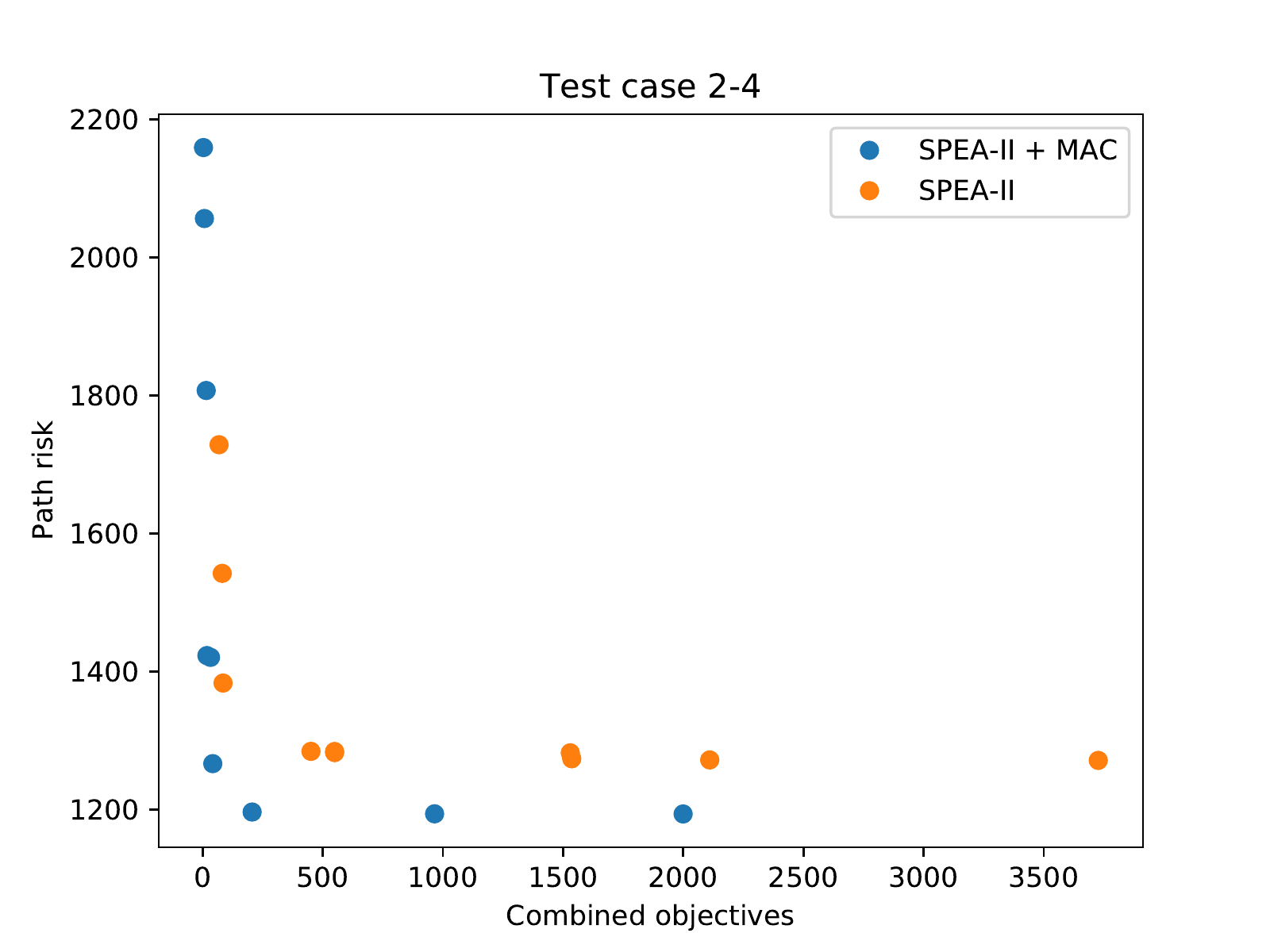}
  \caption{}
\end{subfigure}%
\begin{subfigure}{.35\textwidth}
  \centering
  \includegraphics[width=\linewidth]{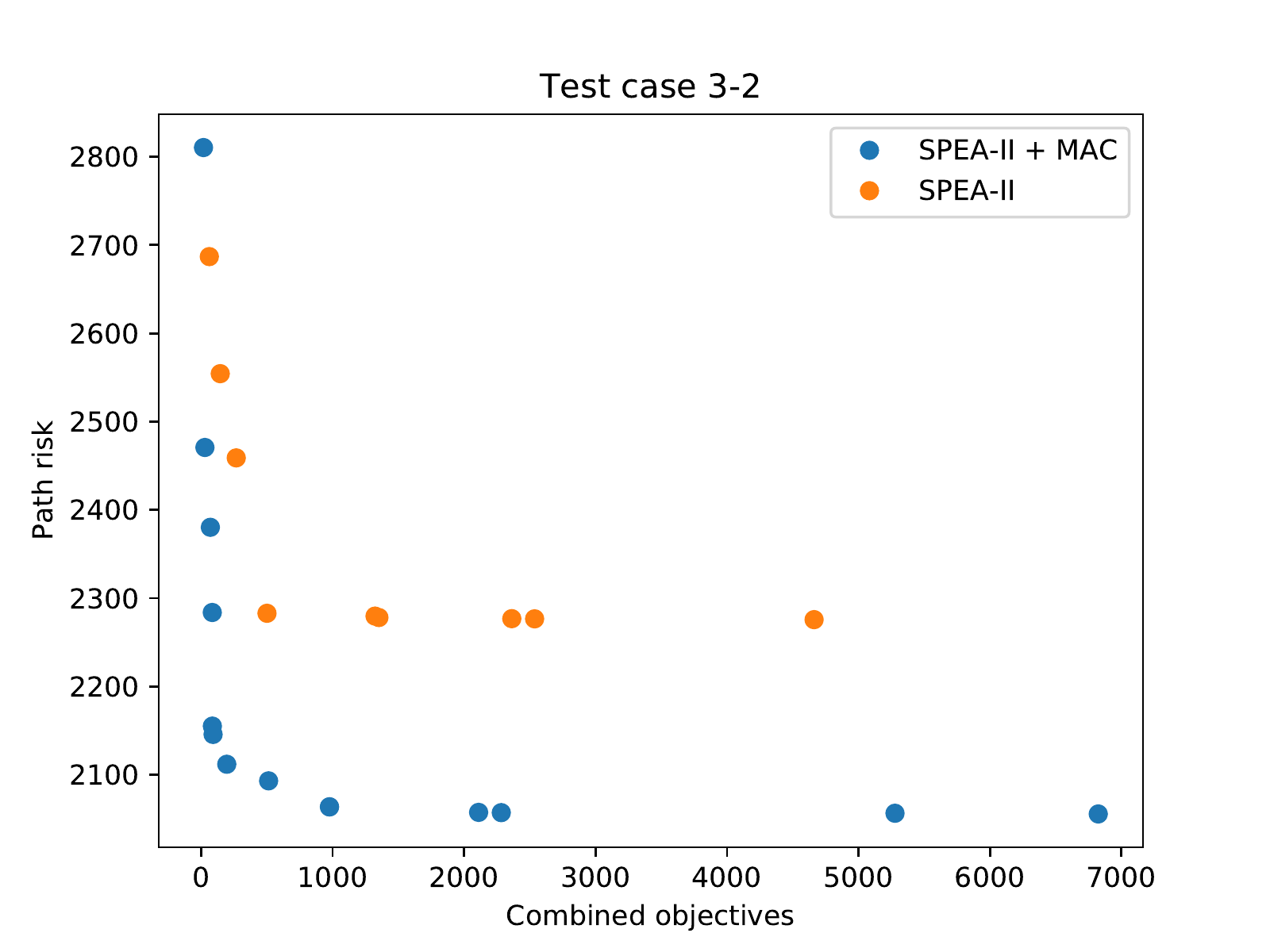}
  \caption{}
\end{subfigure}
\caption{The plot of the non-dominated solutions obtained using applied algorithms with (blue points) and without parameter tuning (red points).}
\label{fig:pareto}
\end{figure*}
\begin{figure*}[h]
\begin{subfigure}{.35\textwidth}
  \centering
  \includegraphics[width=\linewidth]{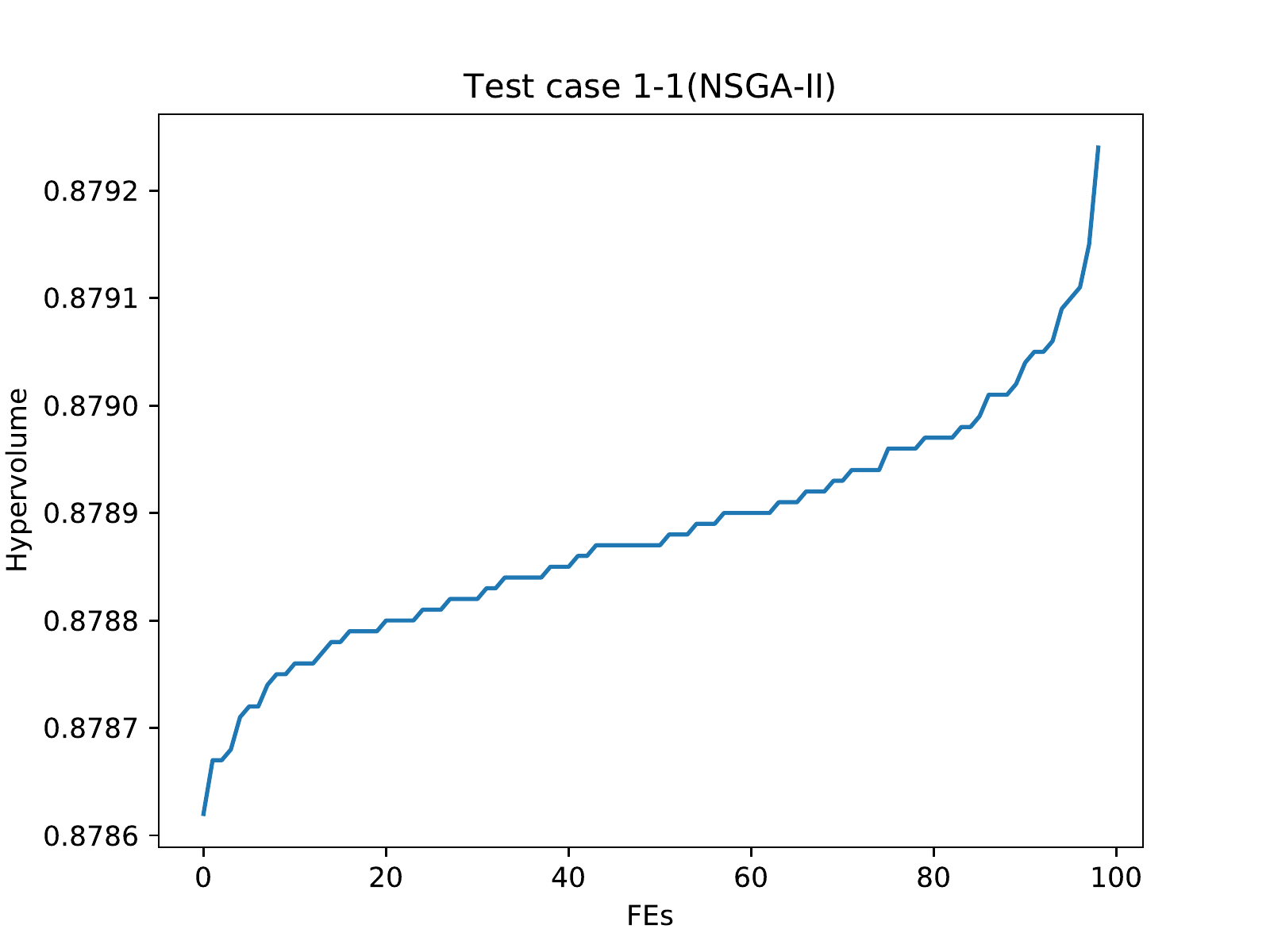}
  \caption{}
\end{subfigure}%
\begin{subfigure}{.35\textwidth}
  \centering
  \includegraphics[width=\linewidth]{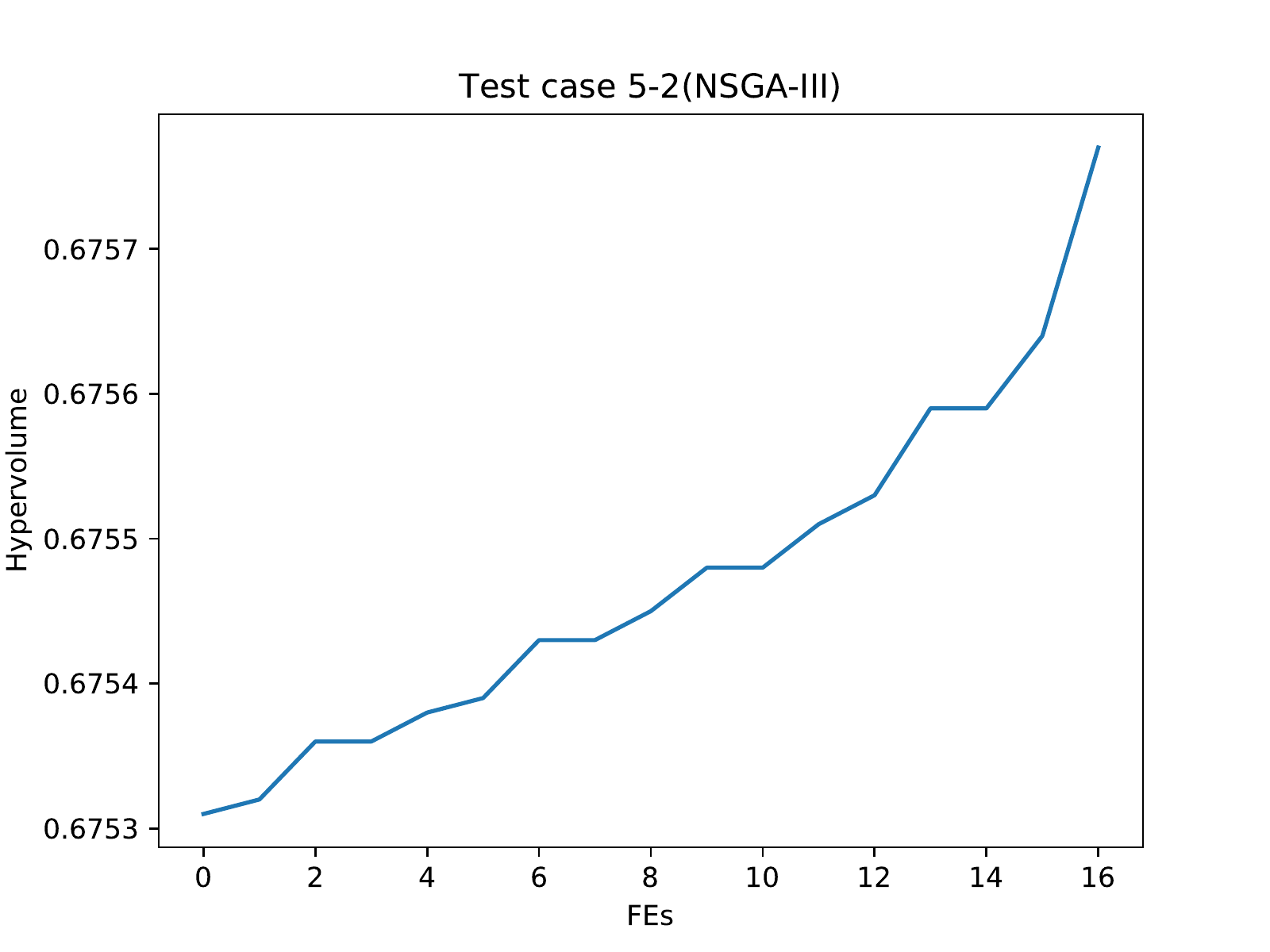}
  \caption{}
\end{subfigure}
\begin{subfigure}{.35\textwidth}
  \centering
  \includegraphics[width=\linewidth]{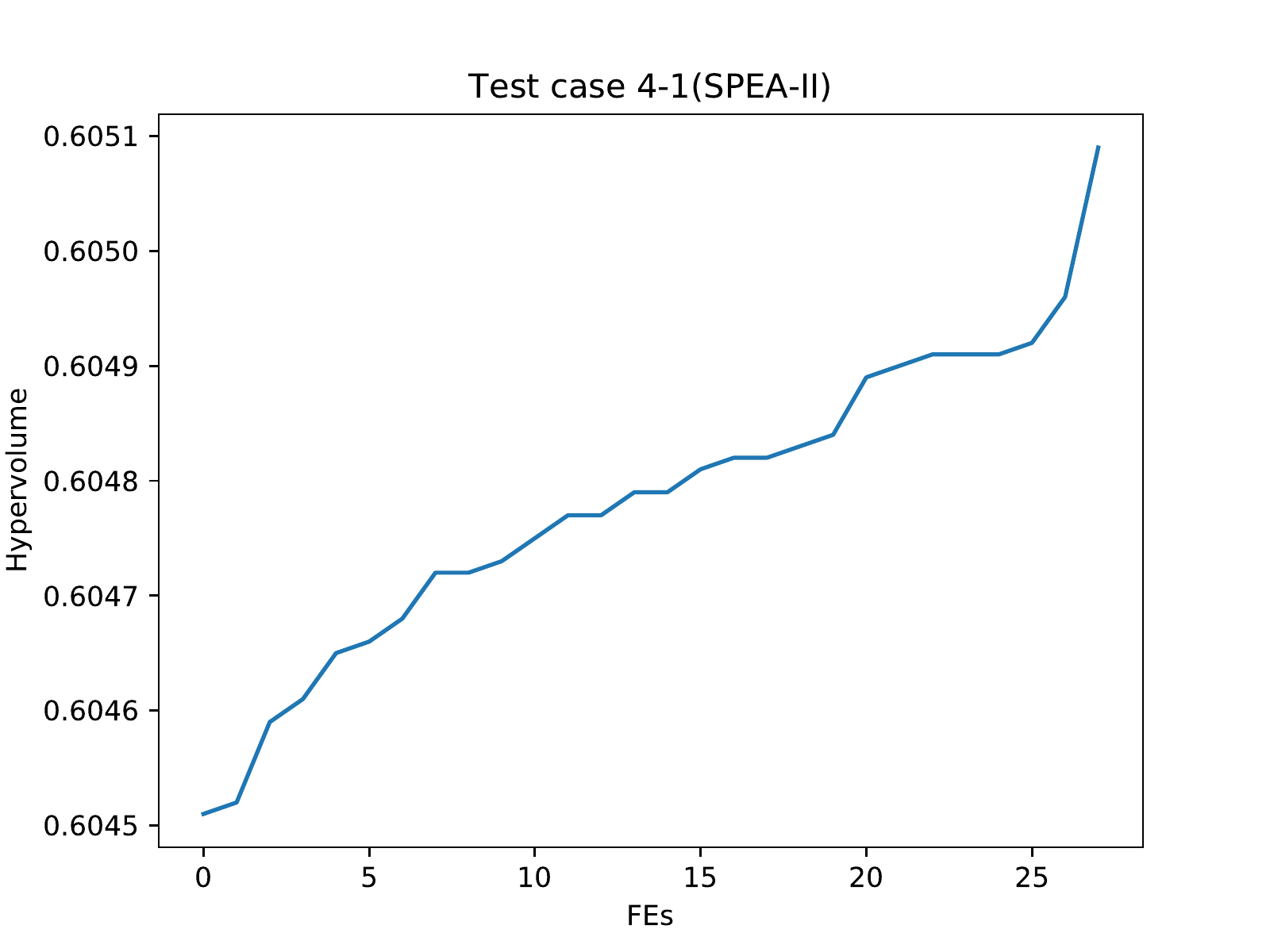}
  \caption{}
\end{subfigure}%

\caption{The plot of convergence rate of HV for different test cases and algorithms}
\label{fig:HVS}
\end{figure*}

Figure \ref{fig:pareto} displays the Pareto front sets of combined objectives \textit{vs} path risk for NSGA-II, NSGA-III, and SPEA-II with \textit{MAC} (blue points) and without it (red points), respectively. As can be seen, the obtained results show the superiority of \textit{MAC} in finding the best results with a higher number of non-dominated solutions. Furthermore, Figure \ref{fig:HVS} illustrates the convergence rate of \textit{HV} value for different test cases. These figures clarify the impact of applying \textit{MAC} for improving the quality of the results.


\begin{table*}
\scriptsize
\begin{center}
\caption{The percentage of improvement in Hypervolume indicator for different test cases.}\label{table:5}
\begin{tabular}{c c c c c c c}
\hline
\rule{0pt}{8pt}
&\multicolumn{3}{c}{Algorithms + \textit{MAC}}&\multicolumn{3}{c}{Algorithms}\\
\cline{2-7}
\rule{0pt}{8pt}
Dataset &  SPEA-II & NSGA-II &NSGA-III & SPEA-II & NSGA-II &NSGA-III \\
\hline
T1-1 &  100.00\% & 99.96\% & 100.00\% & 95.25\% & 94.67\% & 92.56\% \\
T1-2 &  97.19\% & 100.00\% & 97.10\% & 94.34\% & 93.84\% & 92.60\% \\
T1-3 &  100.00\% & 99.05\% & 97.98\% & 95.60\% & 95.23\% & 96.48\% \\
T1-4 &  99.18\% & 99.88\% & 98.10\% & 95.77\% & 96.94\% & 100.00\% \\
\hline
T2-1 &  98.67\% & 98.51\% & 100.00\% & 99.30\% & 94.64\% & 92.69\% \\
T2-2 &  98.30\% & 100.00\% & 99.44\% & 95.40\% & 96.13\% & 93.26\% \\
T2-3 &  99.56\% & 98.89\% & 100.00\% & 96.26\% & 94.00\% & 93.74\% \\
T2-4 &  98.47\% & 100.00\% & 99.93\% & 96.83\% & 94.34\% & 94.64\% \\
\hline
T3-1 &  99.16\% & 100.00\% & 99.55\% & 95.50\% & 94.63\% & 96.08\% \\
T3-2 &  98.30\% & 100.00\% & 99.82\% & 95.57\% & 97.10\% & 94.88\% \\
T3-3 &  98.97\% & 99.88\% & 100.00\% & 96.13\% & 95.31\% & 97.26\% \\
T3-4 &  96.99\% & 98.82\% & 100.00\% & 95.14\% & 94.23\% & 95.88\% \\
\hline
T4-1 &  100.00\% & 98.49\% & 99.57\% & 94.93\% & 95.02\% & 94.01\% \\
T4-2 &  98.71\% & 99.20\% & 100.00\% & 95.10\% & 96.58\% & 95.31\% \\
T4-3 &  99.53\% & 99.78\% & 100.00\% & 95.33\% & 95.86\% & 95.80\% \\
T4-4 &  98.39\% & 100.00\% & 99.01\% & 97.60\% & 94.58\% & 94.65\% \\
\hline
T5-1 &  97.54\% & 100.00\% & 99.02\% & 95.09\% & 95.73\% & 96.85\% \\
T5-2 &  96.08\% & 97.86\% & 100.00\% & 94.14\% & 93.16\% & 94.42\% \\
T5-3 &  98.57\% & 100.00\% & 98.53\% & 97.12\% & 94.50\% & 94.33\% \\
T5-4 &  98.90\% & 100.00\% & 99.89\% & 95.27\% & 94.95\% & 94.65\% \\
\hline
\end{tabular}
\label{Tab:table1}
\end{center}
\end{table*}




\section{Conclusions and future perspectives}\label{Conc}
This study presents a new mathematical formulation for a multi-objective UAV path planning problem. For this purpose, the path length, energy consumption, and path risk are introduced and considered as objective functions. The novelty of this study is the possibility of bypassing obstacles or adjusting the altitude for flying above them. Besides, customized search operators are suggested for solving the problem. The experiments are accomplished using five randomly generated test cases by applying several EMO algorithms. The results are reported with and without parameter tuning. The obtained results show the effect of applying automatic configuration for parameter tuning and also the superiority of the NSGA-II algorithm.
\par
For future research, based on the dynamics of the considered UAV, the problem could be studied in a dynamic environment comprised of moving and static obstacles especially for 3D real urban environment. Developing other solution algorithms to obtain better results in less computational time could be of great interest.

\section*{Acknowledgment}
This work is part of a research project funded by the French \textit{Agence Nationale de la Recherche} under grant number ANR-16-SEBM-0004. Also, the authors would like to acknowledge heir indebtedness to \textit{Hojjat Rakhshani} for his guidance and expert advice which have been invaluable throughout all stages of the work.

\bibliographystyle{IEEEtran}
\bibliography{wcci}

\begin{thebibliography}{10}
\providecommand{\url}[1]{#1}
\csname url@samestyle\endcsname
\providecommand{\newblock}{\relax}
\providecommand{\bibinfo}[2]{#2}
\providecommand{\BIBentrySTDinterwordspacing}{\spaceskip=0pt\relax}
\providecommand{\BIBentryALTinterwordstretchfactor}{4}
\providecommand{\BIBentryALTinterwordspacing}{\spaceskip=\fontdimen2\font plus
\BIBentryALTinterwordstretchfactor\fontdimen3\font minus
  \fontdimen4\font\relax}
\providecommand{\BIBforeignlanguage}[2]{{%
\expandafter\ifx\csname l@#1\endcsname\relax
\typeout{** WARNING: IEEEtran.bst: No hyphenation pattern has been}%
\typeout{** loaded for the language `#1'. Using the pattern for}%
\typeout{** the default language instead.}%
\else
\language=\csname l@#1\endcsname
\fi
#2}}
\providecommand{\BIBdecl}{\relax}
\BIBdecl

\bibitem{golabi2017edge}
M.~Golabi, S.~M. Shavarani, and G.~Izbirak, ``An edge-based stochastic facility
  location problem in uav-supported humanitarian relief logistics: a case study
  of tehran earthquake,'' \emph{Natural Hazards}, vol.~87, no.~3, pp.
  1545--1565, 2017.

\bibitem{shavarani2019congested}
S.~M. Shavarani, S.~Mosallaeipour, M.~Golabi, and G.~{\.I}zbirak, ``A congested
  capacitated multi-level fuzzy facility location problem: An efficient drone
  delivery system,'' \emph{Computers \& Operations Research}, vol. 108, pp.
  57--68, 2019.

\bibitem{dorling2016vehicle}
K.~Dorling, J.~Heinrichs, G.~G. Messier, and S.~Magierowski, ``Vehicle routing
  problems for drone delivery,'' \emph{IEEE Transactions on Systems, Man, and
  Cybernetics: Systems}, vol.~47, no.~1, pp. 70--85, 2016.

\bibitem{ghambari2018comparative}
S.~Ghambari, J.~Lepagnot, L.~Jourdan, and L.~Idoumghar, ``A comparative study
  of meta-heuristic algorithms for solving uav path planning,'' in \emph{2018
  IEEE Symposium Series on Computational Intelligence (SSCI)}.\hskip 1em plus
  0.5em minus 0.4em\relax IEEE, 2018, pp. 174--181.

\bibitem{samar2012optimal}
R.~Samar and W.~A. Kamal, ``Optimal path computation for autonomous aerial
  vehicles,'' \emph{Cognitive Computation}, vol.~4, no.~4, pp. 515--525, 2012.

\bibitem{bekhti2016path}
M.~Bekhti, M.~Abdennebi, N.~Achir, and K.~Boussetta, ``Path planning of
  unmanned aerial vehicles with terrestrial wireless network tracking,'' in
  \emph{2016 Wireless Days (WD)}.\hskip 1em plus 0.5em minus 0.4em\relax IEEE,
  2016, pp. 1--6.

\bibitem{nikolos2003evolutionary}
I.~K. Nikolos, K.~P. Valavanis, N.~C. Tsourveloudis, and A.~N. Kostaras,
  ``Evolutionary algorithm based offline/online path planner for uav
  navigation,'' \emph{IEEE Transactions on Systems, Man, and Cybernetics, Part
  B (Cybernetics)}, vol.~33, no.~6, pp. 898--912, 2003.

\bibitem{peng2012intelligent}
X.~Peng and D.~Xu, ``Intelligent online path planning for uavs in adversarial
  environments,'' \emph{International Journal of Advanced Robotic Systems},
  vol.~9, no.~1, p.~3, 2012.

\bibitem{kavraki1994probabilistic}
L.~Kavraki, P.~Svestka, and M.~H. Overmars, \emph{Probabilistic roadmaps for
  path planning in high-dimensional configuration spaces}.\hskip 1em plus 0.5em
  minus 0.4em\relax Unknown Publisher, 1994, vol. 1994.

\bibitem{lavalle2001randomized}
S.~M. LaValle and J.~J. Kuffner~Jr, ``Randomized kinodynamic planning,''
  \emph{The international journal of robotics research}, vol.~20, no.~5, pp.
  378--400, 2001.

\bibitem{gammell2014informed}
J.~D. Gammell, S.~S. Srinivasa, and T.~D. Barfoot, ``Informed rrt*: Optimal
  sampling-based path planning focused via direct sampling of an admissible
  ellipsoidal heuristic,'' in \emph{2014 IEEE/RSJ International Conference on
  Intelligent Robots and Systems}.\hskip 1em plus 0.5em minus 0.4em\relax IEEE,
  2014, pp. 2997--3004.

\bibitem{zhao2018survey}
Y.~Zhao, Z.~Zheng, and Y.~Liu, ``Survey on computational-intelligence-based uav
  path planning,'' \emph{Knowledge-Based Systems}, vol. 158, pp. 54--64, 2018.

\bibitem{paredes2017study}
J.~A. Paredes, C.~Saito, M.~Abarca, and F.~Cuellar, ``Study of effects of
  high-altitude environments on multicopter and fixed-wing uavs' energy
  consumption and flight time,'' in \emph{2017 13th IEEE Conference on
  Automation Science and Engineering (CASE)}.\hskip 1em plus 0.5em minus
  0.4em\relax IEEE, 2017, pp. 1645--1650.

\bibitem{li2012efficient}
G.~Li, A.~Yamashita, H.~Asama, and Y.~Tamura, ``An efficient improved
  artificial potential field based regression search method for robot path
  planning,'' in \emph{2012 IEEE International Conference on Mechatronics and
  Automation}.\hskip 1em plus 0.5em minus 0.4em\relax IEEE, 2012, pp.
  1227--1232.

\bibitem{zitzler1998evolutionary}
E.~Zitzler and L.~Thiele, ``An evolutionary algorithm for multiobjective
  optimization: The strength pareto approach,'' \emph{TIK-report}, vol.~43,
  1998.

\bibitem{zitzler2001spea2}
E.~Zitzler, M.~Laumanns, and L.~Thiele, ``Spea2: Improving the strength pareto
  evolutionary algorithm,'' \emph{TIK-report}, vol. 103, 2001.

\bibitem{deb2000fast}
K.~Deb, S.~Agrawal, A.~Pratap, and T.~Meyarivan, ``A fast elitist non-dominated
  sorting genetic algorithm for multi-objective optimization: Nsga-ii,'' in
  \emph{International conference on parallel problem solving from
  nature}.\hskip 1em plus 0.5em minus 0.4em\relax Springer, 2000, pp. 849--858.

\bibitem{deb2013evolutionary}
K.~Deb and H.~Jain, ``An evolutionary many-objective optimization algorithm
  using reference-point-based nondominated sorting approach, part i: solving
  problems with box constraints,'' \emph{IEEE Transactions on Evolutionary
  Computation}, vol.~18, no.~4, pp. 577--601, 2013.

\bibitem{rakhshani2019mac}
H.~Rakhshani, L.~Idoumghar, J.~Lepagnot, and M.~Br{\'e}villiers, ``Mac:
  Many-objective automatic algorithm configuration,'' in \emph{International
  Conference on Evolutionary Multi-Criterion Optimization}.\hskip 1em plus
  0.5em minus 0.4em\relax Springer, 2019, pp. 241--253.

\bibitem{hajela1992genetic}
P.~Hajela and C.-Y. Lin, ``Genetic search strategies in multicriterion optimal
  design,'' \emph{Structural optimization}, vol.~4, no.~2, pp. 99--107, 1992.

\bibitem{zitzler1999multiobjective}
E.~Zitzler and L.~Thiele, ``Multiobjective evolutionary algorithms: a
  comparative case study and the strength pareto approach,'' \emph{IEEE
  transactions on Evolutionary Computation}, vol.~3, no.~4, pp. 257--271, 1999.

\end{thebibliography}

\end{document}